\documentclass[submission,copyright,creativecommons]{eptcs}
\providecommand{\event}{ICLP 2019} 
\usepackage{breakurl}             
\usepackage{underscore}           

\usepackage{mathptmx}

\usepackage{url}  
\usepackage{graphicx}  
\usepackage{xspace}
\usepackage{xcolor}
\usepackage{enumitem}
\usepackage[mathscr]{euscript}
\usepackage{comment}

\def\verb#1{\hbox{\tt #1}\xspace}
\def\class#1{{\hbox{{\sc #1\/}}}}

\def\alm{$\mathscr{ALM}$\xspace}
\def\restKB{{\sc RestKB}\xspace}
\def\corealm{{\sc CoreALMlib}\xspace}

\newtheorem{example}{Example}
\newtheorem{definition}{Definition}

\title{{\sc RestKB}: A Library of Commonsense Knowledge about Dining at a Restaurant}
\author{Daniela Inclezan
\institute{Department of Computer Science and Software Engineering\\
Miami University\\ Oxford, OH, USA}
\email{inclezd@miamioh.edu}
}
\def\titlerunning{{\sc RestKB}: A Library of Commonsense Knowledge about Dining at a Restaurant}
\def\authorrunning{D. Inclezan}
\begin{document}
\maketitle

\begin{abstract}
This paper presents a library of commonsense knowledge, \restKB, developed in modular action language \alm
and containing background knowledge relevant to the understanding of restaurant narratives,
including stories that describe exceptions to the normal unfolding of such scenarios.
We highlight features that KR languages must possess in order to be able to express pertinent knowledge,
and expand action language \alm as needed.
We show that encoding the knowledge base in \alm
facilitates its piecewise construction and testing, 
and improves the generality and 
quality of the captured information, in comparison to an initial ASP encoding.
The knowledge base was used in a system for reasoning
about stereotypical activities, evaluated on the restaurant domain.
\end{abstract}

\section{Introduction}
\label{sec:intro}
In this paper, we introduce a library of general commonsense knowledge related to the restaurant domain, \restKB, which  
is a result of our research on understanding narratives about stereotypical activities, including stories that
describe exceptional scenarios. In previous work \cite{zi17,izbi18,zbi18}, 
we introduced a methodology for reasoning about such stories
and demonstrated the advantages of using ASP \cite{gl91} in implementing it.
The knowledge in \restKB was also originally written in ASP, 
but this encoding soon became difficult to manage and test. 
The next alternative was an action language, i.e., a high-level logic programming language
dedicated to the representation of knowledge about actions and their effects \cite{gl93}.
Particularly appealing was a \emph{modular} action language, \alm \cite{ig16}, 
that allows for the structuring and reuse of knowledge,
and facilitates the gradual development and testing of knowledge modules.
\alm's semantics is described in terms of a translation into
ASP\{f\} \cite{bal13}, an extension of ASP with non-Herbrand functions, 
which means that an \alm knowledge base can be seamlessly integrated into our ASP-based methodology
for reasoning about stories involving stereotypical activities. 
However, \alm was not immediately suitable for our purpose, as its
syntax did not provide means for specifying non-deterministic effects of actions, which was necessary 
for reasoning about exceptional scenarios that require diagnosis.
To better manage the growing amount of encoded knowledge, we 
decided to expand \alm with the missing feature, which was 
facilitated by the close connection with ASP\{f\}.

Background knowledge bases are important for solving a variety of reasoning tasks,
but they are even more important when dealing with narratives about stereotypical activities
(e.g., dining at a restaurant, going to the doctor, etc.).
Stories about stereotypical activities omit many more details about actions that take place,
compared to other texts. These details are assumed to be filled in by the reader,
based on a common understanding of how such activities should unfold. For instance,
in Example \ref{ex:normal} below, the fact that Nicole paid for her meal 
is not explicitly mentioned but assumed to have happened.

\begin{example}[Normal Scenario, adapted from Mueller's work \cite{m07}]
\label{ex:normal}
Nicole went to a vegetarian restaurant. She ordered a lentil soup. The waitress put the soup
on the table. Nicole enjoyed the soup. She left the restaurant.
\end{example} 

\noindent
Most importantly, our methodology is able to understand and explain exception scenarios like the one 
in Example \ref{ex:diag}, which could not be processed by previous approaches \cite{m07}.
A suitable knowledge base should contain information that supports this type of reasoning.

\begin{example}[Scenario Requiring Diagnosis]
\label{ex:diag}
Nicole went to a vegetarian restaurant. She ordered lentil soup. The waitress brought her a miso soup instead. 
{\em (Possible explanations include that the waitress or cook misunderstood the order.)}
\end{example}
 
In particular, the knowledge in \restKB supports reasoning about (a) normal restaurant scenarios; 
(b) scenarios in which something goes wrong (e.g., wrong dish/bill is brought to the customer); 
(c) scenarios in which the activity has to be suddenly stopped (e.g., the customer receives an urgent call and leaves); and
(d) scenarios in which some of the customer's subgoals are serendipitously achieved by someone else's actions
(e.g., someone pays for the customer's bill). 
{\em Exceptional scenarios} of types (b)-(d) cannot be handled in a scalable way by
previous approaches, including Mueller's system.

\smallskip
We focus here on a knowledge base about the restaurant domain because a large number of
actions occurring in dining episodes are general enough to be encountered in other domains
(e.g., enter, go, request). We envision the modules created for the restaurant domain to be
the foundation for a commonsense knowledge base created on top of linguistic resources, such as the 
verb ontology VerbNet \cite{KipperPhd05,verbnet.2006}, 
and applicable to the task of general natural language understanding,
as outlined by Lierler \emph{et al.} \cite{LierlerIG17}. Lierler \emph{et al.} suggest
coupling state-of-the-art natural language processing tools and resources with knowledge representation and
reasoning (KR) techniques to demonstrate a deep understanding of the meaning of 
texts written in natural language. 

The contributions of this paper are as follows:
\begin{itemize}[leftmargin=*,noitemsep,topsep=0pt]
\item We demonstrate the benefits of using a higher-level logic programming language (an action language, specifically
a modular one) to the creation of a restaurant library; 
\item We discuss key KR principles that we applied in creating the \restKB, relevant to
the creation of other libraries;  
\item We present the \alm library \restKB; and
\item We demonstrate the use of \restKB.
\end{itemize}


In the remainder of the paper, we discuss related work and language \alm.
We highlight desired features for high-level logic programming languages to support 
reasoning about (exceptional) stories involving stereotypical activities,
and expand \alm as needed. 
We present \restKB and the key principles employed in creating it. 
We exemplify the use of the library on the restaurant domain, starting from natural language text.


\section{Related Work}
\label{sec:relwork}
Mueller \cite{m07} studied restaurant texts and presented a system that could
process stories about {\em normal} scenarios with a reasonable accuracy. 
His work was a continuation of research by Shank and Abelson \cite{sa77}
and represented a substantial improvement in terms of system capabilities 
compared to system SAM (Script Applier Mechanism) 
\cite{c78}, due to the use of logic programming for reasoning purposes.
Unfortunately, Mueller's background knowledge base
is proprietary and thus not available. Additionally, Mueller's approach cannot handle
exception scenarios of types (b)-(d), like the one in Example \ref{ex:diag}, because it relies on the use of fixed scripts.
This implies that his knowledge base lacks information needed to reason about, or explain,
untypical scenarios.

A previous \alm core library exists, \corealm \cite{di16}, derived from the Component Library \cite{bpc01} 
written in the language KM \cite{cp04}. 
However, in our work with restaurant narratives we discovered that the collections of fluents and axioms
in these two libraries are not rich enough to be useful for a deep understanding of restaurant narratives.
Actions are denoted in natural language by action verbs. 
There are linguistic resources like VerbNet \cite{KipperPhd05} that classify verbs and attempt to provide semantics
for their meaning. However, these are only informal semantics (i.e., annotations) and are not useful in
building reasoning systems.


A modular action language with similar goals to \alm's is MAD \cite{lr06}. A MAD library of core concepts
exists \cite{thesiser08} but it requires substantial expansion to be applicable to the restaurant domain and
it cannot be directly integrated in our ASP-based system for reasoning
about stereotypical activities.
Moreover, the reuse mechanism of MAD has the potential of requiring a higher number of
modules than its \alm correspondent and a deeper module hierarchy \cite{di15}.
These are less desirable features from a software engineering point of view.


\section{Action Language \alm}
\label{sec:alm}

Language \alm \cite{ig16} is a recent KR language for modeling dynamic domains.
It is a {\em modular} action language that provides means
for the structuring of knowledge into {\em reusable} classes and modules, which facilitates 
the knowledge engineering task, and the piecewise construction and testing of libraries and system descriptions. An \alm module is a formal description of a specific piece of knowledge packaged as a unit, and consists of declarations of classes, functions, and axioms.

In \alm, the goal is to represent {\em classes} (sorts) of actions and objects in general terms 
(e.g, a $move$ action class) instead of particular actions and objects (e.g., Nicole going to the table).
Axioms written about a class are general and apply to all specific instances of the class.
Moreover, classes have attributes that are optional instead of fixed parameters
(e.g., a $move$ action class has attributes $origin$ and $dest$ but any of these can be
omitted from the definition of an instance of $move$). 
This facilitates the mapping of natural language stories into an \alm logic form. 

Here is an example of an \alm action class declaration:

$
\begin{array}{l}
move :: {\bf actions}\\
\ \ \ {\bf attributes}\\
\ \ \ \ \ \ actor : agents\\
\ \ \ \ \ \ origin, dest : points
\end{array}
$

\noindent
This declares $move$ as an action class with three attributes: $actor$ of sort $agents$, and
$origin$ and $dest$ of sort $points$. Fluents (and statics) are declared using a syntax similar to that of 
mathematical functions, for instance

$at : things \rightarrow points$

\noindent
says that $at$, describing the location of things, maps $things$ into $points$.
Fluents, statics, and attributes are all (possibly partial) functions. 
Axioms describe the effects of actions and conditions for their execution. 
As an example, the axiom below states that the direct effect of
the occurrence of an action of class $move$ is that its actor will be at the destination.

$
\begin{array}{lrlrl}
occurs(X) & {\bf causes} & at(A) = D & {\bf if} & instance(X, move),\ actor(X) = A,\ dest(X) = D.
\end{array}
$

\noindent
The next axiom states that an instance of $move$ cannot be executed if its actor is already
at the destination.

$
\begin{array}{llll}
{\bf impossible} & occurs(X) & {\bf if} & instance(X, move),\ actor(X) = A,\ dest(X) = D,\ at(A) = D.
\end{array}
$

An example of a definition of an action class instance is shown below. It represents Nicole's action
of going to a vegetarian restaurant (note that the $origin$ is not specified):

$
\begin{array}{l}
e_1 \ {\bf in} \ move\\
\ \ \ \ \ \ actor = ``Nicole\mbox{''}\\
\ \ \ \ \ \ dest = ``a\ vegetarian\ restaurant\mbox{''}
\end{array}
$

The semantics of \alm is given via a translation into ASP\{f\} \cite{bal13},
e.g., the translation of the two \alm axioms above looks as follows:

$
\begin{array}{lll}
at(A, I+1) = D & \leftarrow & occurs(X, I),\ instance(X, move),\ actor(X) = A,\ dest(X) = D.
\end{array}
$

$
\begin{array}{lll}
\neg occurs(X, I) & \leftarrow & instance(X, move),\ actor(X) = A,\ dest(X) = D,\ at(A, I) = D.
\end{array}
$

\noindent
where $I$ ranges over a new sort $step$, $A$ ranges over $agents$ and $D$ over $points$.
The translation also includes pre-defined rules like the Inertia Axioms for inertial fluents.



\section{Desired Features in KR Languages}
\label{sec:features}

As we were building our knowledge base for the restaurant domain, 
we determined a set of features that KR languages (especially action languages)
should possess to support the understanding of stories about stereotypical activities.
We list here the identified features and justify their need.

\smallskip
\noindent
{\bf 1. An elegant solution to allowing optional attributes of actions (or action classes) instead of fixed parameters.}
In natural language, the arguments of verbs (also called thematic/semantic roles in linguistic terms) 
are often optional. For instance, we could encounter the verb ``leave'' either with a specified origin as in 
``She left the restaurant'' (see Example \ref{ex:normal}) or without an explicit origin as in 
``She left.'' An action language in which actions/ action classes have optional attributes
would streamline the translation from natural language into the vocabulary of a knowledge base.

\smallskip
\noindent
{\bf 2. An ability to create concise representations for a large number of actions/ English verbs.}
Many actions are relevant to the understanding of stories about stereotypical activities,
especially if multiple actors are involved as is the case with the restaurant domain. 
For restaurant scenarios we identified sixteen relevant actions in previous research \cite{izbi18}
(thirteen in Mueller's work). 
To facilitate the knowledge engineering task, action languages should allow a compact 
and speedy representation of a large number of
actions that map into an even larger number of English verbs. 
Ideally, action languages would possess means for mimicking the
way verbs are defined in natural language in terms of other, 
more basic verbs (e.g., {\em carry} defined as a special case of {\em move}),
and the grouping of verbs via synonymy.

\smallskip
\noindent
{\bf 3. Means for representing partial fluents/ functions.}
In many dynamic domains including restaurant dining, 
items are created that did not exist before (e.g., a dish is prepared), or items are consumed. 
Fluents associated with these objects, such as location, 
should be undefined in states prior to the object coming into existence or after the object ceases to exist as is 
(e.g., it does not make sense to talk about the location of a dish before it is prepared). 

\smallskip
\noindent
{\bf 4. Means for encoding non-deterministic effects of actions.}
This is particularly important for scenarios with exceptions, especially the ones that require explanations. 
For instance, in the story in Example \ref{ex:diag}, the possible explanation that the waitress misunderstood the order 
requires communication actions to be encoded as possibly having non-deterministic results under certain conditions.
This would allow concluding that the waitress either understands the order to be for a lentil or a miso soup
if there is interference.

\subsection{\alm and the Identified Features}
The syntax and semantics of modular action language \alm cover features {\bf 1--3} above. 
\alm attributes (e.g., $origin$, $dest$) are optional in the sense
that they may or may not be instantiated in an instance definition; 
thus feature {\bf 1} is satisfied. With respect to feature {\bf 2}, 
\alm has means for declaring classes of actions, and for declaring classes 
in terms of priorly defined classes. It can also structure information into modules of knowledge, 
consisting of declarations of sorts, functions, and axioms,
which can be reused (imported) when building new modules, and can be independently tested. 
As for feature {\bf 3}, \alm fluents are by default partial functions, 
unless the keyword {\bf total} precedes their declarations. 
In terms of semantics, functions are translated into non-Herbrand functions of ASP\{f\}, 
which can be partial. 
Additionally, for each function $f$, \alm provides a pre-defined function $dom_f$ 
that is true if $f$ is defined for a given set of parameters and false otherwise. 
This allows specifying that a fluent becomes undefined as a result of some action occurring, 
as in the examples for feature {\bf 3} above.

Feature {\bf 4} required some changes to the syntax of \alm. 
To illustrate this, we consider a $request$ action class with attributes 
$actor$, $item$ requested, and $recipient$ of the request. 
To handle miscommunication scenarios like the one in Example \ref{ex:diag}, 
the library should contain some rule stating that, if an $interference$ action occurs simultaneously,
then a request has a non-deterministic effect in terms of what the $actor$ understands to be
the requested $item$.
Specifically, we would like to encode this knowledge in \alm via rules like:
\begin{equation}
\label{eq1}
\begin{array}{lrl}
occurs(X) & {\bf causes} & informed(R, T, A) \\
          &     {\bf if} & instance(X, request), \ recipient(X) = R,\ item(X) = T,\\
					&              & actor(X) = A,\ \neg occurs(interference).
\end{array}
\end{equation}
\begin{equation}
\label{eq2}
\begin{array}{lrl}
occurs(X) & {\bf causes} & 1 \{informed(R, T1, A)\ :\ instance(T1, things), T1 \neq T\}\ 1 \\
          &     {\bf if} & instance(X, request),\ recipient(X) = R,\ item(X) = T,\\
          &              & actor(X) = A,\ occurs(interference).
\end{array}
\end{equation}
\noindent
Such rules are not in the syntax of \alm because $
occurs$ expressions are not allowed in the body of dynamic causal laws, 
though they are allowed in executability conditions (see (\ref{eq1})), 
and choice elements\footnote{Intuitively, choice elements are syntactic instruments that allow 
a compact way of describing multiple possibilities/models.} are not allowed in the heads of rules (see (\ref{eq2})).
Dynamic causal laws of \alm have the syntax:

$occurs(a)\ {\bf causes}\ f (\bar{x} ) = o\ {\bf if}\ instance(a, c),\ cond$

\noindent
where $a$ and $o$ are variables or constants, $f$ is a basic (i.e., inertial) fluent, 
$c$ is the class $actions$ or a subclass of it, and $cond$ is a collection of literals. 
The law says that an occurrence of an action $a$ of the class $c$ in a state satisfying property 
$cond$ causes the value of $f (\bar{x} )$ to become $o$ in any resulting state.

\smallskip
We change the syntax of \alm to now allow $cond$ to contain expressions of the form 
$occurs(t)$ or $\neg occurs(t)$ where $t$ is a variable or an object constant of the sort $actions$. 
We also allow the head of dynamic rules $f (\bar{x} ) = o$ to be replaced by a 
choice element in {\sc gringo} syntax (as in (\ref{eq2})), where the choice must concern a basic (i.e., inertial) fluent.
This implies no changes to the semantics of \alm (i.e., the translation into ASP\{f\}
and definition of a transition) 
because the existing definition of the translation 
already specifies how $occurs$ expressions should be processed, 
given that they may appear in the body of executability conditions, and
{\sc gringo}-style choice elements are part of the language of ASP\{f\}. 

An additional refinement of \alm is related to one of the KR principles discussed in the next section. 
The original version of \alm does not allow overloaded attributes, 
meaning the same attribute name declared in different classes. 
However, it would be convenient to reuse the same attribute name (e.g., $actor$) 
in several action classes. We relax the initial constraint by adding the following 
definition of a {\em valid attribute} and requiring that all attributes in a system description
must be valid.

\begin{definition}[Valid Attribute]
\label{def:attr}
An attribute $a$ is {\em valid} if 

\noindent
(1) there is at most one declaration of $a$ per class and 

\noindent
(2) each axiom containing an attribute literal of the form $a(X, \bar{x}) = o$ or $a(X, \bar{x}) \neq o$, 
where $X$ is a variable, ${\bar x}$ is a (possibly empty) collection of variables/ constants, 
and $o$ is a variable/ constant, 
must also contain an atom of the form $instance(X, c)$ where $c$ is a class. 
\end{definition}

\noindent
With the new requirement, if both action classes $move$ and $request$ have an attribute $actor$, 
an axiom of the form:

$head \ \ {\bf if} \ \ instance(X, request), actor(X) = o.$

\noindent
would be valid, but the same rule without an atom of the type $instance(X, s)$ would not.



\section{The \restKB Library}
\label{sec:library}

In building our library for the restaurant domain, we were guided by the following key {\em KR principles}
that we believe to be important for the creation of other libraries, possibly in other KR or action languages as well:
\begin{enumerate}[leftmargin=*,noitemsep,topsep=0pt]
\item[I.] {\bf A shallow action class inheritance hierarchy and module dependency hierarchy.}

Research on object-oriented software maintainability indicates that an inheritance depth of three is optimal
\cite{dbmrw96}.

\item[II.] {\bf Introduce new attributes as low as possible in the action class hierarchy.}

In contrast, in the Component Library \cite{bpc01} the superclass for all action classes
contains all possible attributes (thematic roles) that any action (verb) may have.
As needed, action classes are then specified not to contain a given attribute or 
to have a restricted attribute range.
We believe that this approach makes it difficult for people using the library to identify the names and range
of attributes of a given action class, which may require looking into the top class.
With the principle we propose, attribute information will reside locally with the action class 
and will be easier to find.

\item[III.] {\bf Reduce the overall number of attribute names.}

The idea would be to reuse attribute names whenever possible. For instance, 
$actor$ would be a common attribute for multiple action classes. Note that the original version of \alm
required different names for attributes of different action classes (e.g., 
$mover$, $grasper$). To accommodate this principle, we relaxed the definition
of a valid attribute declaration in the previous section, under the conditions specified in Definition \ref{def:attr}.

\item[IV.] {\bf Restrict attributes of an action class to necessary ones.}

This will improve the readability of action classes. 
By necessary we mean attributes that are commonly used in conjunction 
with a given action class or appear in axioms. For instance, we recommend not introducing an 
attribute $instrument$ in action class $prepare$ (some food) if $instrument$ does not appear in any of the axioms.

\item[V.] {\bf Place opposite concepts in the same module.}
 
This applies to action classes with opposite effects (e.g., $enter$ and $leave$),
fluents that are (almost) opposites (e.g., $standing$ and $sitting$), and opposite axioms.

\item[VI.] {\bf Limit the number of conditions in the body of dynamic causal laws.}

Instead of increasing the number of conditions in dynamic causal laws, 
delegate as many conditions as possible to executability conditions for a separation of purposes of axiom types.
\end{enumerate}

\begin{figure}[!htbp]
\centering
\fbox{\includegraphics[width=0.85\textwidth]
{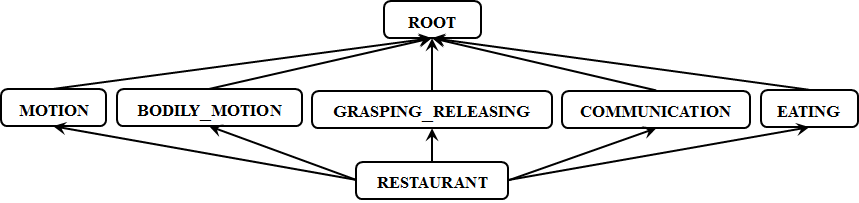}}
\caption{Module Dependency Hierarchy in \restKB}
\label{fig:modulehier}
\end{figure}

\smallskip
We applied these KR principles in building \restKB.
The library declares 30 classes out of which nineteen are action classes; seventeen inertial fluents; and 95 axioms. 
Classes (and the pertinent fluents and axioms) are grouped into 
seven modules based on their common theme.
Module names and the dependencies between them are shown in Figure \ref{fig:modulehier}.
All of the modules in \restKB except {\sc restaurant} 
are general enough to be useful in modeling other domains.
Specifically, in preliminary work on the ``going to the doctor'' stereotypical activity, 
we determined that modules {\sc motion} and {\sc communication} would be relevant 
and could be imported from this library (e.g., the patient {\em goes} from the reception area to the exam room;
the patient {\em communicates} with the receptionist/ doctor). 
The classes and inertial fluents in each module are shown in Table \ref{table:contents}.
Note that $universe$ and $actions$ are built-in classes of \alm, and 
additional {\em defined} fluents of \alm (i.e., defined in terms of other fluents) are present in some of the modules.

\begin{table}[!hbt]
\caption{Contents of \restKB modules}
\label{table:contents}
\small
\begin{minipage}{\textwidth}
\begin{tabular}{l l l l}
\hline\hline
{\bf Module} & {\bf Contents} & \\
\hline
{\sc root} & Classes: & $agents$, $things$, $points$, $areas$  & subclasses of $universe$\\
	         &          & $persons$                              & subclass of $agents$\\
	         & Fluents: & $at$, $in$ &\\
\hline
{\sc motion} & Classes: & $move$, $enter$, $leave$, $lead\_to$ & subclasses of $actions$\\
	           & Fluents: & $open$ &\\
\hline
{\sc bodily\_motion} & Classes: & $sit$, $stand\_up$ & subclasses of $actions$ \\
                     & Fluents: & $sitting$, $standing$, $seating\_exists$ & \\
\hline
{\sc grasping\_releasing} & Classes: & $grasp$, $release$ & subclasses of $actions$ \\
                          &          & $put\_on$              & subclass of $release$ \\
                          & Fluents: & $holding$, $can\_reach$ & \\ 
\hline
{\sc communication} & Classes: & $communicate$, $interference$ & subclasses of $actions$\\
                    &          & $greet$, $request$ & subclasses of $communicate$\\
                    & Fluents: & $greeted\_by$, $informed$ &  \\
\hline
{\sc eating} & Classes: & $foods$          & subclass of $things$ \\
             &          & $eat$, $prepare$ & subclasses of $actions$ \\ 
						 & Fluents: & $satiated$, $food\_prepared\_by$ & \\ 
\hline
{\sc restaurant} & Classes: & $customers$, $waiters$, $cooks$ & subclasses of $persons$  \\
                 &          & $bills$                         & subclass of $things$ \\
								 &          & $restaurants$                   & subclass of $areas$ \\
								 &          & $read\_menu$, $pay$, $becomes\_unavailable$ & subclasses of $actions$\\
								 &          & $order$                         & subclass of $request$ \\
								 & Fluents: & $available$, $served$, $has\_read\_menu$,  & \\
								 &          & $bill\_generated\_for$, $paid$ &\\
\hline\hline
\end{tabular}
\end{minipage}
\end{table}
\normalsize

Modules contain between 2 and 9 classes out of which 0--5 are action classes;
1--5 inertial fluents; and 0--50 axioms. 
The average numbers per module are: 4 classes out of which 3 action classes; 2 inertial fluents; and 13 axioms. 
The depth of the class hierarchy is three.

We compared \restKB with our initial ASP encoding and 
found the following advantages for the \alm library:
\begin{itemize}[leftmargin=*,noitemsep,topsep=0pt]
\item {\em More concise.}
About 20 of the axioms in the original ASP encoding
did not need to be represented in the \alm version. 
This is a result of \alm's power to define subclass relations 
(e.g., $order$ is a subclass of $request$) leading to axiom reuse,
and to declare {\em functional} fluents (i.e., fluents with a unique value, like $at$, for which specifying a non-boolean range in \alm automatically 
implies that the value is unique).
\item {\em Higher quality.} 
\alm axioms follow principle VI listed at the beginning of this section, while the initial ASP encoding did not.
Also, \restKB contains 10\% new relevant axioms 
that were missing from the ASP version, as well as two new action classes and 
two new fluents. Some of these were relevant specifically to the
restaurant domain, but others were added to increase the generality of the solution
and allow modules to be suitable for other domains. 
\item {\em More manageable and easier to test,} as the knowledge is divided into modules.
\end{itemize}

The library (available at \url{https://tinyurl.com/yan3qam5}) was tested via integration in a system for reasoning about stereotypical activities \cite{izbi18}, where the ASP translation of \restKB
replaced the initial ASP knowledge base. 
A collection of 20 restaurant stories was used in the evaluation, and the \restKB-based system 
performed as well as the original. 

\section{Tools for Library Use}

To facilitate the use of the \restKB library in natural language understanding tasks, we created
two tables that can be searched for relevant \restKB knowledge.
A first table connects action classes and fluents 
to word senses from WordNet \cite{f05}, 
as done in previous work on commonsense libraries \cite{di16}. 
WordNet is a large lexical database for the English language. 
It groups nouns, verbs, adjectives, and adverbs into ``sets of cognitive synonyms, each expressing a distinct concept.'' 
In our first table, word senses from WordNet 3.1\footnote{\url{http://wordnetweb.princeton.edu/perl/webwn}}
represent the search keys. 
In Table \ref{table:searchWordNet}, 
we show parts of this table for the action classes 
and fluents from module {\sc motion} of \restKB.
Note, for example, that WordNet senses \verb{go\#1}, \verb{locomote\#1}, and \verb{travel\#1}
are synonyms (i.e., they have the same definition), 
and therefore searching by any of these terms
leads to the same \restKB action class, $move$.

\begin{table}[!hbt]
\caption{Searching library contents by WordNet 3.1 verbs (motion verbs)}
\label{table:searchWordNet}
\small
\begin{minipage}{\textwidth}
\begin{tabular}{l|l}
\hline\hline
{\bf WordNet 3.1 Verb\ \ [WordNet Synonyms]} & {\bf \restKB} \\
 & {\bf Action Class}\\
\hline
\verb{go\#1} \ \ [\verb{locomote\#1, travel\#1}] & $move$ \\
\ \ \ \ \ \ Definition: change location; move, travel,  or proceed, also metaphorically & \\
\hline
\verb{enter\#1} \ \ [\verb{come in\#1,get in\#1,get into\#2,go in\#1,go into\#1,move into\#1}]\ \ \ \ \ \  & $enter$\\
\ \ \ \ \ \ Definition: to come or go into & \\
\hline
\verb{leave\#1} \ \ [\verb{go away\#2, go forth\#1}] & $leave$\\
\ \ \ \ \ \ Definition: go away from a place & \\
\hline
\verb{leave\#5} \ \ [\verb{exit\#1, get out\#1, go out\#1, go out\#1 }] & $leave$\\
\ \ \ \ \ \ Definition: move out of or depart from & \\
\hline
\verb{lead\#1} \ \ [\verb{conduct\#4, direct\#5, guide\#2, take\#3}] & $lead\_to$\\
\ \ \ \ \ \ Definition: take somebody somewhere & \\
\hline\hline
\end{tabular}
\end{minipage}
\end{table}
\normalsize

Additionally, we connect action classes to verb classes
from the verb ontology VerbNet \cite{KipperPhd05} and predicates from PropBank \cite{propbank}, following the approach suggested by Lierler \emph{et al.} \cite{LierlerIG17}.
VerbNet\footnote{\url{https://uvi.colorado.edu/}} is a verb lexicon that categorizes English verbs 
based on their syntactico-semantic behavior in sentences.
VerbNet classes may contain several verbs. For instance, the verb class \class{escape-51.1-1} contains among others verb \verb{go}; 
one of its subclasses, \class{escape-51.1-1-2}, contains verb \verb{enter}.
Note that \restKB also contains an action class $enter$ that is a subclass of $move$ (the matching action class for verb \verb{go} of VerbNet).
As seen in Table~\ref{table:searchVerbNet}, not all English 
verbs have a clear mapping into a VerbNet class (e.g., the verb \verb{release}
as in ``letting go of an object'' corresponding to our action class $release$ may be mapped into class \class{let-64.2}, but this
is not completely clear; the verb \verb{interfere} corresponding to our action class $interference$ is not linked to any VerbNet class yet.)
PropBank~\cite{propbank}\footnote{\url{http://verbs.colorado.edu/~mpalmer/projects/ace.html}} is a linguistic resource that provides information on the semantic roles (i.e., arguments) associated with different verbs (i.e., predicate-argument relations). 
The PropBank predicates listed in Table~\ref{table:searchVerbNet} have sense information associated with them coming from Ontonotes Sense Groupings.\footnote{\url{http://clear.colorado.edu/compsem/index.php?page=lexicalresources&sub=ontonotes}}
For instance, the suffix \verb{01} in \verb{go.01} indicates that Ontonotes associates this PropBank predicate with sense~1 of verb \verb{go}. 
This second table can be searched by VerbNet verb class
or PropBank predicate. 


\begin{table}[!hbt]
\caption{Searching library contents by VerbNet verb classes
and PropBank predicates}
\label{table:searchVerbNet}
\small
\begin{minipage}{\textwidth}
\begin{tabular}{l l l l}
\hline\hline
{\bf VerbNet Verb Class} & {\bf PropBank Predicate} & {\bf \restKB Action Class} \ \ &
{\bf \restKB Module}\\
\hline
\class{escape-51.1-1} & \verb{go.01}, \verb{go.02}& move & \sc{motion}\\
\class{escape-51.1-1-2} & \verb{enter.01} & enter & \\
\class{escape-51.1-1-1} & \verb{leave.01}, \verb{leave.04} \ \ \ \ & leave & \\
\class{accompany-51.7} & \verb{lead.01} & lead\_to & \\
\hline
\class{assuming\_position-50} \ \ \ \ & \verb{sit.01} & sit & \sc{bodily\_motion}\\
\class{assuming\_position-50} & \verb{stand.01} & stand\_up & \\
\hline
\class{hold-15.1-1} & \verb{grasp.01} & grasp & \sc{grasping\_releasing}\\
\class{let-64.2} (?) & \verb{release.01} & release & \\
\class{put-9.1-2} & \verb{put.01} & put & \\
\hline
\class{transfer\_mesg-37.1.1} & \verb{communicate.01} & communicate & \sc{communication}\\
--                & \verb{interfere.01} & interference & \\
\class{judgment-33.1-1} & \verb{greet.01} & greet & \\
\class{beg-58.2} & \verb{request.01} & request & \\
\hline
\class{eat-39.1-1} & \verb{eat.01} & eat & \sc{eating}\\
\class{preparing-26.3-1} & \verb{prepare.01} & prepare & \\
\hline
\class{pay-68.1} & \verb{pay.01} & pay & \sc{restaurant}\\
\class{get-13.5-1} & \verb{order.02} & order & \\
\hline\hline
\end{tabular}
\end{minipage}
\end{table}
\normalsize


\section{Application: Restaurant Stories}

We used \restKB as the background commonsense knowledge base for a question answering system \cite{izbi18}
dedicated to restaurant scenarios of the type defined in the Section~\ref{sec:intro}. 
Our system can answer questions of the following types:
\begin{itemize}[noitemsep,topsep=0pt]
\item $query\_yes\_no(A)$ -- Did action $A$ occur?
\item $query\_when(A)$ -- When did action $A$ occur?
\item $query\_where(P,A)$ -- Where was person $P$ when action $A$ happened?
\item $query\_who(A)$ -- Who performed action $A$?
\item $query\_who\_whom(A)$ -- Who performed action $A$ and to whom?
\item $query\_what(F,A)$ -- What was the value of fluent $F$ when action $A$ happened?
\item $query\_goal(P,A)$ -- What goal was $P$ trying to achieve when action $A$ happened?
\item $query\_intended(P,A)$ -- What was $P$'s intended activity when action $A$ happened?
\end{itemize}
where $A$ is a physical action.
In the spirit of Muller's work \cite{m07}, 
we recently expanded our system with an ASP module that can 
generate a number of queries for each input text,
such that the answers to these queries are not explicitly stated in the text \cite{zbi18}.

In previous work \cite{izbi18}, we exemplified how the information in the input text can be connected
to axioms in the knowledge base, which was originally written in ASP. 
In this section, we will show that encoding the knowledge base in \alm actually 
facilitates the process of connecting the natural language in the input text to axioms in the \alm knowledge base
{\em in a way that can be automated}. 
As before, we use the approach outlined by Lierler {\em et al.} \cite{LierlerIG17},
which relies on a variety of state-of-the-art NLP tools. We illustrate this process on Example \ref{ex:normal}.

\begin{figure}[!htbp]
\centering
\begin{tabular}{cc}
(a) {\sc LTH} output & (b) {\sc CoreNLP} coreference output\\
		\begin{minipage}{.53\textwidth}
			\fbox{\includegraphics[width=0.97\textwidth]{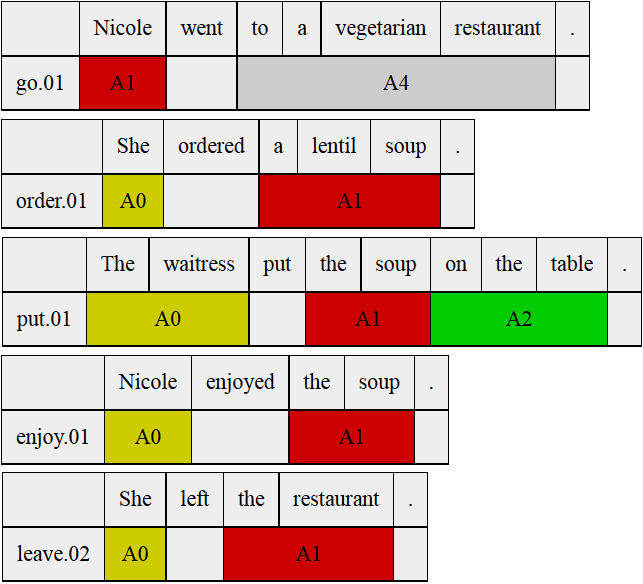}}			 
		\end{minipage}
		&
		\begin{minipage}{.46\textwidth}
			\fbox{\includegraphics[width=0.90\textwidth]{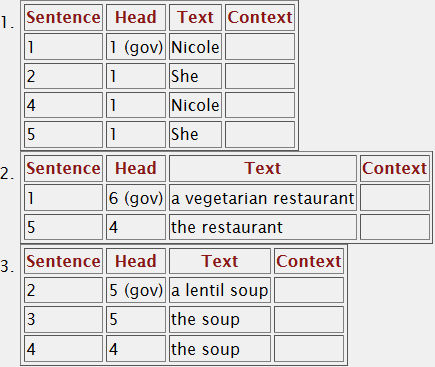}}			
		\end{minipage}
\end{tabular}
\caption{Output from state-of-the-art NLP tools for the text in Example \ref{ex:normal}}
\label{fig:NLPoutput}
\end{figure}

As a first step, the input text would be fed into a number of NLP tools that perform parsing, semantic role labeling, mention detection, and coreference resolution.
We learned that we can obtain more complete information by using several tools instead of just one \cite{LierlerIG17}.
The output of the LTH semantic role labeler,\footnote{\url{http://barbar.cs.lth.se:8081/parse}}
shown in Figure \ref{fig:NLPoutput}(a),
indicates what verbs were identified in the text, including their Ontonotes Sense Groupings (e.g., \verb{go.01}) 
and their semantic roles/ arguments (e.g., goer), which are annotated by LTH with PropBank labels (e.g., A1).
Note that the same semantic role label can mean different things for different verbs. 
The PropBank semantic role labels and their meanings
for the relevant verbs in Example \ref{ex:normal} are shown in Table~\ref{table:PropBankroles}.
The output produced by the CoreNLP system,\footnote{\url{http://nlp.stanford.edu:8080/corenlp/}} 
and shown in Figure~\ref{fig:NLPoutput}(b), 
includes mention detection and coreference resolution.
This output correctly indicates, for example, that pronoun ``she'' in sentences 2 and 5 
refers to ``Nicole,'' who is mentioned in sentences 1 and 4.

\begin{table}[!hbt]
\caption{PropBank semantic roles}
\label{table:PropBankroles}
\small
\begin{minipage}{\textwidth}
\begin{tabular}{l | l | l| l}
\hline\hline
\verb{go.01} & \verb{order.02} & \verb{put.01} & \verb{leave.01}\\
(motion) & (request to be delivered) & (location) & (depart)\\
\hline
{\bf A1: entity in motion/goer}		& {\bf A0: orderer}				& {\bf A0: putter}		& {\bf A0: entity in motion}\\
A2: extent									& {\bf A1: thing ordered}	& {\bf A1: thing put} &{\bf A1: starting point}\\
{\bf A3: start point}							& A2: benefactive, ordered-for\ \ \ & {\bf A2: where put} \ \ \ & A2: destination\\
{\bf A4: end point, end state of A1} \ \ \ & {\bf A3: source} & & \\
\hline\hline
\end{tabular}
\end{minipage}
(Semantic roles that have a correspondent in the matching action classes of \restKB are shown in bold.)
\end{table}
\normalsize

Given this input, our goal is to produce a logic form representation of the input text 
that allows us to use axioms in \restKB to reason and answer queries
about the text. This logic form should contain 
facts defining objects named in the text as instances of relevant \restKB sorts, and observations about the occurrences of action instances explicitly mentioned in the text.
In other words, we need to connect the output of the NLP tools shown above
to the vocabulary of \restKB. In Lierler {\em et al.}'s approach, this is done via an intermediary step 
that constructs a so called Discourse Representation Structure (DRS) \cite{kampreyle93}. 
Due to space limitations, we skip this step here (see \cite{LierlerIG17,izbi18} for details).

To achieve our goal, Table \ref{table:searchVerbNet} (which  
enables searching \restKB by PropBank verbs)
needs to be expanded with a matching of PropBank semantic roles to attributes of action classes of \restKB.
Table \ref{table:PropBankrestKB} indicates the mappings necessary to process the text in Example~\ref{ex:normal}.
For instance, semantic role labels A1, A3, A4 of the verb \verb{go.01} match the attributes
$actor$, $origin$, and $dest$, respectively, of action class $move$ of module {\sc motion} in \restKB.
Note that this matching is not perfect, as some PropBank verbs have additional semantic roles
with respect to their \restKB correspondent (e.g., \verb{order.02} in PropBank has the semantic role
``A2: benefactive, ordered-for'' that does not have a correspondent in the action class $order$ of \restKB). 

\begin{table}[!hbt]
\caption{Mapping of PropBank semantic roles to \restKB attributes}
\label{table:PropBankrestKB}
\small
\begin{minipage}{\textwidth}
\begin{tabular}{l l | l l}
\hline\hline
{\bf PropBank Semantic Role} & {\bf \restKB Attribute}\ \ \ & {\bf PropBank Semantic Role} & {\bf \restKB Attribute}\\
\hline
\verb{go.01} & $\mathbf{move}$ & \verb{order.02} & {\bf order}\\
\ \ \ A1: entity in motion/goer & \ \ \ $actor$ & \ \ \ A0: orderer  & \ \ \ $actor$\\
\ \ \ A3: start point           & \ \ \ $origin$ & \ \ \ A1: thing ordered & \ \ \ $item$\\
\ \ \ A4: end point, end state of A1 \ \ \ & \ \ \ $dest$ & \ \ \ A2: source & \ \ \ $recipient$ \\
\hline
\verb{put.01} & {\bf put\_on} & \verb{leave.01} & {\bf leave} \\
\ \ \ A0: putter      & \ \ \ $actor$ & \ \ \ A0: entity in motion & \ \ \ $actor$\\
\ \ \ A1: thing put   & \ \ \ $object$ & \ \ \ A1: starting point  & \ \ \ $origin$\\
\ \ \ A2: where put   & \ \ \ $on$ & &\\
\hline\hline
\end{tabular}
\end{minipage}
\end{table}
\normalsize

Using the output of LTH and CoreNLP shown in Figure \ref{fig:NLPoutput} 
together with the matching provided in Table \ref{table:PropBankrestKB},
the following action instances could be automatically produced:

\noindent
$
\begin{array}{l}
\ \ \ \ \ \ \ \ \ e_1\ {\bf in}\ move \\
\ \ \ \ \ \ \ \ \ \ \ \ actor = ``Nicole\mbox{''} \\
\ \ \ \ \ \ \ \ \ \ \ \ dest = ``a\ vegetarian\ restaurant\mbox{''}
\end{array}
$

\noindent
$
\begin{array}{l}
\ \ \ \ \ \ \ \ \ e_2\ {\bf in}\ order\\
\ \ \ \ \ \ \ \ \ \ \ \ actor = ``Nicole\mbox{''} \\
\ \ \ \ \ \ \ \ \ \ \ \ item = ``a\ lentil\ soup\mbox{''} 
\end{array}
$
																									
\noindent
$
\begin{array}{l}
\ \ \ \ \ \ \ \ \ e_3\ {\bf in}\ put\_on\\
\ \ \ \ \ \ \ \ \ \ \ \ actor = ``the\ waitress\mbox{''} \\
\ \ \ \ \ \ \ \ \ \ \ \ object = ``a\ lentil\ soup\mbox{''} \\
\ \ \ \ \ \ \ \ \ \ \ \ on = ``the\ table\mbox{''} 
\end{array}
$

\noindent
$
\begin{array}{l}
\ \ \ \ \ \ \ \ \ e_4\ {\bf in}\ leave\\
\ \ \ \ \ \ \ \ \ \ \ \ actor = ``Nicole\mbox{''} 
\end{array}
$

\noindent
For instance, to produce $e_1$ the system would analyze 
the LTH output shown in Figure \ref{fig:NLPoutput}(a) 
and identify \verb{go.01} as the verb in the first sentence.
It would then search Table \ref{table:searchVerbNet} with key \verb{go.01}
to find the matching action class from \restKB, $move$.
The system would also detect the semantic role labels in the LTH output, A1 and A4, and the text entities they correspond to, ``Nicole'' and ``a vegetarian restaurant.'' By searching Table
\ref{table:PropBankrestKB} with keys A1 and A4, the system would identify the attributes
$actor$ and $dest$ of $move$, respectively, and connect them to
the appropriate text entities. 

After generating action instances, the system would then produce facts about the occurrence of these action instances, as required by the logic form encoding of the text
expected by our methodology for processing restaurant stories, i.e.,
$\{st\_hpd(e_1, true, 0), st\_hpd(e_2, true, 1), st\_hpd(e_3, true, 2)$, $st\_hpd(e_4, true, 3)\}$, saying that $e_1$ was the first event explicitly mentioned to have occurred in the story, $e_2$ was the second one, etc.
The logic form will in turn trigger the appropriate axioms from \restKB, for instance axioms (\ref{eq1}) and \ref{eq2}) 
about $request$ actions,
as $e_2$ is an instance of action class $order$, which is a subclass of $request$.

In contrast, when using the original ASP background knowledge base, 
the process of producing a logic form could not be easily automated; instead, one rule had to be written for
each action class of the knowledge base, as shown in the example below \cite{izbi18}:

\noindent
$
\begin{array}{lll}
st\_hpd(go(Actor,Dest),true,S) & \leftarrow & event(Ev), eventType(Ev,go\_01),\\
& & eventArgs(Ev,a1,EActor), property(EActor,Actor), \\
& & eventArgs(Ev,a4,EDest), property(EDest,Dest).
\end{array}
$

Being able to automate the process of translating the input text into a logic form 
according to Lierler {\em et al.}'s proof-of-concept is a substantial improvement.
The automation of this process is an important research question in itself with its own hurdles.
The output of NLP tools is not always accurate or complete, and thus requires added logic for remediation.
As an example, sense \verb{order.02} (``request to be delivered'') 
is a better match for verb ``order'' as used in Example \ref{ex:normal}
than the sense \verb{order.01} (``impelled action'') identified by the LTH system.
Also, sort information for the entities appearing in a text can be inferred from the semantic roles
they instantiate in verbs, but usually this information is not as specific as the information produced by hand.


\section{Conclusions}
\label{sec:conclusions}

We have presented an \alm library of commonsense knowledge about restaurants, \restKB.
We have identified features that KR languages should possess to support reasoning about narratives
describing stereotypical activities, including exceptional scenarios,
and expanded \alm as needed. We have shown that the \alm encoding is more compact, manageable, easier to test,
and of a higher quality than an original ASP encoding.
The knowledge base was integrated in a prototype system for reasoning
about stereotypical activities, evaluated on restaurant stories.

In previous work \cite{LierlerIG17}, we outlined a process for scaling the 
creation of a vast \alm library that can serve as a background knowledge base
for reasoning about a larger number of stereotypical activities.
Our idea is to create \alm representations of the most frequent {\em classes}
of action verbs in English and utilize existing linguistic 
resources in this process (e.g., verb ontology VerbNet).
\alm's features that support the reuse of knowledge will speed up this process
and facilitate the testing of the resulting library.

\nocite{*}
\bibliographystyle{eptcs}
\bibliography{iclp2019}

\begin{thebibliography}{10}
\providecommand{\bibitemdeclare}[2]{}
\providecommand{\surnamestart}{}
\providecommand{\surnameend}{}
\providecommand{\urlprefix}{Available at }
\providecommand{\url}[1]{\texttt{#1}}
\providecommand{\href}[2]{\texttt{#2}}
\providecommand{\urlalt}[2]{\href{#1}{#2}}
\providecommand{\doi}[1]{doi:\urlalt{http://dx.doi.org/#1}{#1}}
\providecommand{\bibinfo}[2]{#2}

\bibitemdeclare{inproceedings}{bal07b}
\bibitem{bal07b}
\bibinfo{author}{Marcello \surnamestart Balduccini\surnameend}
  (\bibinfo{year}{2007}): \emph{\bibinfo{title}{{CR-MODELS}: {A}n {I}nference
  {E}ngine for {CR-P}rolog}}.
\newblock In: {\sl \bibinfo{booktitle}{Proceedings of LPNMR-07}}, pp.
  \bibinfo{pages}{18--30}, \doi{10.1007/978-3-540-72200-7_4}.

\bibitemdeclare{article}{bal13}
\bibitem{bal13}
\bibinfo{author}{Marcello \surnamestart Balduccini\surnameend}
  (\bibinfo{year}{2013}): \emph{\bibinfo{title}{{ASP} with non-{H}erbrand
  Partial Functions: a Language and System for Practical Use}}.
\newblock {\sl \bibinfo{journal}{Theory and Practice of Logic Programming}}
  \bibinfo{volume}{13}(\bibinfo{number}{4--5}), pp. \bibinfo{pages}{547--561},
  \doi{10.1017/S1471068413000343}.

\bibitemdeclare{article}{bg03d}
\bibitem{bg03d}
\bibinfo{author}{Marcello \surnamestart Balduccini\surnameend} \&
  \bibinfo{author}{Michael \surnamestart Gelfond\surnameend}
  (\bibinfo{year}{2003}): \emph{\bibinfo{title}{Diagnostic {R}easoning with
  {A}-{P}rolog}}.
\newblock {\sl \bibinfo{journal}{Theory and Practice of Logic Programming}}
  \bibinfo{volume}{3}, pp. \bibinfo{pages}{425--461},
  \doi{10.1017/S1471068403001807}.

\bibitemdeclare{inproceedings}{bg03a}
\bibitem{bg03a}
\bibinfo{author}{Marcello \surnamestart Balduccini\surnameend} \&
  \bibinfo{author}{Michael \surnamestart Gelfond\surnameend}
  (\bibinfo{year}{2003}): \emph{\bibinfo{title}{{L}ogic {P}rograms with
  {C}onsistency-{R}estoring {R}ules}}.
\newblock In: {\sl \bibinfo{booktitle}{Proceedings of Commonsense-03}},
  \bibinfo{publisher}{{AAAI} {P}ress}, pp. \bibinfo{pages}{9--18}.

\bibitemdeclare{inproceedings}{bg08}
\bibitem{bg08}
\bibinfo{author}{Marcello \surnamestart Balduccini\surnameend} \&
  \bibinfo{author}{Michael \surnamestart Gelfond\surnameend}
  (\bibinfo{year}{2008}): \emph{\bibinfo{title}{The {AAA} Architecture: An
  Overview}}.
\newblock In: {\sl \bibinfo{booktitle}{Architectures for Intelligent
  Theory-Based Agents, Papers from the 2008 {AAAI} Spring Symposium, 2008}},
  \bibinfo{publisher}{{AAAI} {P}ress}, pp. \bibinfo{pages}{1--6},
  \doi{10.1.1.329.3633}.

\bibitemdeclare{inproceedings}{bpc01}
\bibitem{bpc01}
\bibinfo{author}{Ken \surnamestart Barker\surnameend}, \bibinfo{author}{Bruce
  \surnamestart Porter\surnameend} \& \bibinfo{author}{Peter \surnamestart
  Clark\surnameend} (\bibinfo{year}{2001}): \emph{\bibinfo{title}{A {L}ibrary
  of {G}eneric {C}oncepts for {C}omposing {K}nowledge {B}ases}}.
\newblock In: {\sl \bibinfo{booktitle}{Proceedings of the First International
  Conference on Knowledge Capture}}, \bibinfo{series}{K-CAP '01},
  \bibinfo{publisher}{ACM}, \bibinfo{address}{New York, NY, USA}, pp.
  \bibinfo{pages}{14--21}, \doi{10.1145/500737.500744}.

\bibitemdeclare{techreport}{ccmpst09}
\bibitem{ccmpst09}
\bibinfo{author}{Vinay~K. \surnamestart Chaudhri\surnameend},
  \bibinfo{author}{Peter~E. \surnamestart Clark\surnameend},
  \bibinfo{author}{Sunil \surnamestart Mishra\surnameend},
  \bibinfo{author}{John \surnamestart Pacheco\surnameend},
  \bibinfo{author}{Aaron \surnamestart Spaulding\surnameend} \&
  \bibinfo{author}{Jing \surnamestart Tien\surnameend} (\bibinfo{year}{2009}):
  \emph{\bibinfo{title}{AURA: Capturing Knowledge and Answering Questions on
  Science Textbooks}}.
\newblock \bibinfo{type}{Technical Report}, \bibinfo{institution}{SRI
  International}.

\bibitemdeclare{inproceedings}{cjmpps07}
\bibitem{cjmpps07}
\bibinfo{author}{Vinay~K. \surnamestart Chaudhri\surnameend},
  \bibinfo{author}{Bonnie~E. \surnamestart John\surnameend},
  \bibinfo{author}{Sunil \surnamestart Mishra\surnameend},
  \bibinfo{author}{John \surnamestart Pacheco\surnameend},
  \bibinfo{author}{Bruce \surnamestart Porter\surnameend} \&
  \bibinfo{author}{Aaron \surnamestart Spaulding\surnameend}
  (\bibinfo{year}{2007}): \emph{\bibinfo{title}{Enabling Experts to Build
  Knowledge Bases from Science Textbooks}}.
\newblock In: {\sl \bibinfo{booktitle}{Proceedings of the Fourth International
  Conference on Knowledge Capture}}, \bibinfo{series}{K-CAP '07},
  \bibinfo{publisher}{ACM}, \bibinfo{address}{New York, NY, USA}, pp.
  \bibinfo{pages}{159--166}, \doi{10.1145/1298406.1298435}.

\bibitemdeclare{inproceedings}{ccbchjpsty07}
\bibitem{ccbchjpsty07}
\bibinfo{author}{Peter \surnamestart Clark\surnameend},
  \bibinfo{author}{{Shaw-yi} \surnamestart Chaw\surnameend},
  \bibinfo{author}{Ken \surnamestart Barker\surnameend}, \bibinfo{author}{Vinay
  \surnamestart Chaudhri\surnameend}, \bibinfo{author}{Phil \surnamestart
  Harrison\surnameend}, \bibinfo{author}{Bonnie \surnamestart John\surnameend},
  \bibinfo{author}{Bruce \surnamestart Porter\surnameend},
  \bibinfo{author}{Aaron \surnamestart Spaulding\surnameend},
  \bibinfo{author}{John \surnamestart Thompson\surnameend} \&
  \bibinfo{author}{Peter~Z. \surnamestart Yeh\surnameend}
  (\bibinfo{year}{2007}): \emph{\bibinfo{title}{Capturing and Answering
  Questions Posed to a Knowledge-Based System}}.
\newblock In: {\sl \bibinfo{booktitle}{Proceedings of Fourth International
  Conference on Knowledge Capture}}, \bibinfo{series}{K-CAP '07},
  \bibinfo{publisher}{ACM}, \bibinfo{address}{New York, NY, USA},
  \doi{10.1145/1298406.1298419}.

\bibitemdeclare{misc}{cp04}
\bibitem{cp04}
\bibinfo{author}{Peter~E. \surnamestart Clark\surnameend} \&
  \bibinfo{author}{Bruce \surnamestart Porter\surnameend}
  (\bibinfo{year}{2004}): \emph{\bibinfo{title}{{KM} -- {T}he {K}nowledge
  {M}achine 2.0: {U}sers {M}anual}}.
\newblock \bibinfo{howpublished}{Retrieved from the web page:
  http://www.cs.utexas.edu/users/mfkb/km/userman.pdf}.

\bibitemdeclare{misc}{c78}
\bibitem{c78}
\bibinfo{author}{Richard~E. \surnamestart Cullingford\surnameend}
  (\bibinfo{year}{1978}): \emph{\bibinfo{title}{Script application: Computer
  understanding of newspaper stories}}.
\newblock \bibinfo{howpublished}{Research Report 116}.

\bibitemdeclare{article}{dbmrw96}
\bibitem{dbmrw96}
\bibinfo{author}{John \surnamestart Daly\surnameend}, \bibinfo{author}{Andrew
  \surnamestart Brooks\surnameend}, \bibinfo{author}{James \surnamestart
  Miller\surnameend}, \bibinfo{author}{Marc \surnamestart Roper\surnameend} \&
  \bibinfo{author}{Murray \surnamestart Wood\surnameend}
  (\bibinfo{year}{1996}): \emph{\bibinfo{title}{Evaluating inheritance depth on
  the maintainability of object-oriented software}}.
\newblock {\sl \bibinfo{journal}{Empirical Software Engineering}}
  \bibinfo{volume}{1}(\bibinfo{number}{2}), pp. \bibinfo{pages}{109--132},
  \doi{10.1007/BF00368701}.

\bibitemdeclare{incollection}{el06}
\bibitem{el06}
\bibinfo{author}{Selim \surnamestart Erdo\v{g}an\surnameend} \&
  \bibinfo{author}{Vladimir \surnamestart Lifschitz\surnameend}
  (\bibinfo{year}{2006}): \emph{\bibinfo{title}{Actions as Special Cases}}.
\newblock In: {\sl \bibinfo{booktitle}{Principles of Knowledge Representation
  and Reasoning: Proceedings of the International Conference}},
  \bibinfo{publisher}{AAAI Press}, pp. \bibinfo{pages}{377--387}.

\bibitemdeclare{phdthesis}{thesiser08}
\bibitem{thesiser08}
\bibinfo{author}{Selim~Turhan \surnamestart Erdo\v{g}an\surnameend}
  (\bibinfo{year}{2008}): \emph{\bibinfo{title}{A {L}ibrary of
  {G}eneral-{P}urpose {A}ction {D}escriptions}}.
\newblock Ph.D. thesis, \bibinfo{school}{University of Texas at Austin},
  \bibinfo{address}{Austin, TX, USA}.

\bibitemdeclare{incollection}{f05}
\bibitem{f05}
\bibinfo{author}{Christiane \surnamestart Fellbaum\surnameend}
  (\bibinfo{year}{2005}): \emph{\bibinfo{title}{WordNet and Wordnets}}.
\newblock In \bibinfo{editor}{Keith \surnamestart Brown\surnameend}, editor:
  {\sl \bibinfo{booktitle}{Encyclopedia of Language and Linguistics, Second
  Edition}}, \bibinfo{publisher}{Elsevier}, pp. \bibinfo{pages}{665--670}.

\bibitemdeclare{article}{gl91}
\bibitem{gl91}
\bibinfo{author}{Michael \surnamestart Gelfond\surnameend} \&
  \bibinfo{author}{Vladimir \surnamestart Lifschitz\surnameend}
  (\bibinfo{year}{1991}): \emph{\bibinfo{title}{{C}lassical {N}egation in
  {L}ogic {P}rograms and {D}isjunctive {D}atabases}}.
\newblock {\sl \bibinfo{journal}{New Generation Computing}}
  \bibinfo{volume}{9}(\bibinfo{number}{3/4}), pp. \bibinfo{pages}{365--386},
  \doi{10.1007/BF03037169}.

\bibitemdeclare{article}{gl93}
\bibitem{gl93}
\bibinfo{author}{Michael \surnamestart Gelfond\surnameend} \&
  \bibinfo{author}{Vladimir \surnamestart Lifschitz\surnameend}
  (\bibinfo{year}{1993}): \emph{\bibinfo{title}{{R}epresenting {A}ction and
  {C}hange by {L}ogic {P}rograms}}.
\newblock {\sl \bibinfo{journal}{Journal of Logic Programming}}
  \bibinfo{volume}{17}(\bibinfo{number}{2--4}), pp. \bibinfo{pages}{301--321},
  \doi{10.1016/0743-1066(93)90035-F}.

\bibitemdeclare{article}{gl98}
\bibitem{gl98}
\bibinfo{author}{Michael \surnamestart Gelfond\surnameend} \&
  \bibinfo{author}{Vladimir \surnamestart Lifschitz\surnameend}
  (\bibinfo{year}{1998}): \emph{\bibinfo{title}{{A}ction languages}}.
\newblock {\sl \bibinfo{journal}{Electronic Transactions on AI}}
  \bibinfo{volume}{3}(\bibinfo{number}{16}), pp. \bibinfo{pages}{193--210}.

\bibitemdeclare{inproceedings}{di15}
\bibitem{di15}
\bibinfo{author}{Daniela \surnamestart Inclezan\surnameend}
  (\bibinfo{year}{2015}): \emph{\bibinfo{title}{On the Relationship Between Two
  Modular Action Languages: {A} Translation from {MAD} into {ALM}}}.
\newblock In \bibinfo{editor}{Francesco \surnamestart Calimeri\surnameend},
  \bibinfo{editor}{Giovambattista \surnamestart Ianni\surnameend} \&
  \bibinfo{editor}{Miroslaw \surnamestart Truszczynski\surnameend}, editors:
  {\sl \bibinfo{booktitle}{Logic Programming and Nonmonotonic Reasoning - 13th
  International Conference, {LPNMR} 2015, Lexington, KY, USA, September 27-30,
  2015. Proceedings}}, {\sl \bibinfo{series}{Lecture Notes in Computer
  Science}} \bibinfo{volume}{9345}, \bibinfo{publisher}{Springer}, pp.
  \bibinfo{pages}{411--424}, \doi{10.1007/978-3-319-23264-5_34}.

\bibitemdeclare{article}{di16}
\bibitem{di16}
\bibinfo{author}{Daniela \surnamestart Inclezan\surnameend}
  (\bibinfo{year}{2016}): \emph{\bibinfo{title}{Core{ALM}lib: An {ALM} library
  translated from the Component Library}}.
\newblock {\sl \bibinfo{journal}{Theory and Practice of Logic Programming}}
  \bibinfo{volume}{16}(\bibinfo{number}{5-6}), pp. \bibinfo{pages}{800--816},
  \doi{10.1017/S1471068416000363}.

\bibitemdeclare{article}{ig16}
\bibitem{ig16}
\bibinfo{author}{Daniela \surnamestart Inclezan\surnameend} \&
  \bibinfo{author}{Michael \surnamestart Gelfond\surnameend}
  (\bibinfo{year}{2016}): \emph{\bibinfo{title}{Modular {A}ction {L}anguage
  {ALM}}}.
\newblock {\sl \bibinfo{journal}{Theory and Practice of Logic Programming}}
  \bibinfo{volume}{16}(\bibinfo{number}{2}), pp. \bibinfo{pages}{189–--235},
  \doi{10.1017/S1471068415000095}.

\bibitemdeclare{inproceedings}{izbi18}
\bibitem{izbi18}
\bibinfo{author}{Daniela \surnamestart Inclezan\surnameend},
  \bibinfo{author}{Qinglin \surnamestart Zhang\surnameend},
  \bibinfo{author}{Marcello \surnamestart Balduccini\surnameend} \&
  \bibinfo{author}{Ankush \surnamestart Israney\surnameend}
  (\bibinfo{year}{2018}): \emph{\bibinfo{title}{An {ASP} Methodology for
  Understanding Narratives about Stereotypical Activities}}.
\newblock \bibinfo{volume}{18}, pp. \bibinfo{pages}{535--552},
  \doi{10.1017/S1471068418000121}.

\bibitemdeclare{book}{kampreyle93}
\bibitem{kampreyle93}
\bibinfo{author}{Hans \surnamestart Kamp\surnameend} \& \bibinfo{author}{Uwe
  \surnamestart Reyle\surnameend} (\bibinfo{year}{1993}):
  \emph{\bibinfo{title}{From discourse to logic}}.
\newblock \bibinfo{volume}{1,2}, \bibinfo{publisher}{Kluwer},
  \doi{10.1007/978-94-011-2066-1}.

\bibitemdeclare{inproceedings}{verbnet.2006}
\bibitem{verbnet.2006}
\bibinfo{author}{Karen \surnamestart Kipper\surnameend}, \bibinfo{author}{Anna
  \surnamestart Korhonen\surnameend}, \bibinfo{author}{Neville \surnamestart
  Ryant\surnameend} \& \bibinfo{author}{Martha \surnamestart Palmer\surnameend}
  (\bibinfo{year}{2006}): \emph{\bibinfo{title}{Extending VerbNet with Novel
  Verb Classes}}.
\newblock In: {\sl \bibinfo{booktitle}{Proceedings of the Fifth International
  Conference on Language Resources and Evaluation -- LREC'06}},
  \bibinfo{volume}{2.2}, p.~\bibinfo{pages}{1}.

\bibitemdeclare{phdthesis}{KipperPhd05}
\bibitem{KipperPhd05}
\bibinfo{author}{Karin \surnamestart Kipper-Schuler\surnameend}
  (\bibinfo{year}{2005}): \emph{\bibinfo{title}{VerbNet: A Broad-Coverage,
  Comprehensive Verb Lexicon}}.
\newblock Ph.D. thesis, \bibinfo{school}{University of Pennsylvania}.

\bibitemdeclare{inproceedings}{LierlerIG17}
\bibitem{LierlerIG17}
\bibinfo{author}{Yuliya \surnamestart Lierler\surnameend},
  \bibinfo{author}{Daniela \surnamestart Inclezan\surnameend} \&
  \bibinfo{author}{Michael \surnamestart Gelfond\surnameend}
  (\bibinfo{year}{2017}): \emph{\bibinfo{title}{Action Languages and Question
  Answering}}.
\newblock In: {\sl \bibinfo{booktitle}{{IWCS} 2017 - 12th International
  Conference on Computational Semantics - Short papers, Montpellier, France,
  September 19 - 22, 2017}}.

\bibitemdeclare{inproceedings}{lr06}
\bibitem{lr06}
\bibinfo{author}{Vladimir \surnamestart Lifschitz\surnameend} \&
  \bibinfo{author}{Wanwan \surnamestart Ren\surnameend} (\bibinfo{year}{2006}):
  \emph{\bibinfo{title}{{A} {M}odular {A}ction {D}escription {L}anguage}}.
\newblock \bibinfo{series}{Proceedings of the 21st National Conference on
  Artificial Intelligence (AAAI)}, pp. \bibinfo{pages}{853--859}.

\bibitemdeclare{incollection}{McCHay69}
\bibitem{McCHay69}
\bibinfo{author}{John \surnamestart McCarthy\surnameend} \&
  \bibinfo{author}{Patrick~J. \surnamestart Hayes\surnameend}
  (\bibinfo{year}{1969}): \emph{\bibinfo{title}{{S}ome {P}hilosophical
  {P}roblems from the {S}tandpoint of {A}rtificial {I}ntelligence}}.
\newblock In \bibinfo{editor}{B.~\surnamestart Meltzer\surnameend} \&
  \bibinfo{editor}{D.~\surnamestart Michie\surnameend}, editors: {\sl
  \bibinfo{booktitle}{Machine Intelligence 4}}, \bibinfo{publisher}{Edinburgh
  University Press}, pp. \bibinfo{pages}{463--502},
  \doi{10.1016/B978-0-934613-03-3.50033-7}.

\bibitemdeclare{article}{m07}
\bibitem{m07}
\bibinfo{author}{Erik~T. \surnamestart Mueller\surnameend}
  (\bibinfo{year}{2007}): \emph{\bibinfo{title}{Modelling Space and Time in
  Narratives about Restaurants}}.
\newblock {\sl \bibinfo{journal}{Literary and Linguistic Computing}}
  \bibinfo{volume}{22}(\bibinfo{number}{1}), pp. \bibinfo{pages}{67--84},
  \doi{10.1093/llc/fql014}.

\bibitemdeclare{article}{propbank}
\bibitem{propbank}
\bibinfo{author}{Martha \surnamestart Palmer\surnameend},
  \bibinfo{author}{Daniel \surnamestart Gildea\surnameend} \&
  \bibinfo{author}{Paul \surnamestart Kingsbury\surnameend}
  (\bibinfo{year}{2005}): \emph{\bibinfo{title}{The Proposition Bank: An
  Annotated Corpus of Semantic Roles}}.
\newblock {\sl \bibinfo{journal}{Computational Linguistics}}
  \bibinfo{volume}{31}(\bibinfo{number}{1}), pp. \bibinfo{pages}{71--106},
  \doi{10.1162/0891201053630264}.

\bibitemdeclare{book}{sa77}
\bibitem{sa77}
\bibinfo{author}{R.~C. \surnamestart Schank\surnameend} \&
  \bibinfo{author}{R.~P. \surnamestart Abelson\surnameend}
  (\bibinfo{year}{1977}): \emph{\bibinfo{title}{Scripts, Plans, Goals, and
  Understanding: An Inquiry into Human Knowledge Structures}}.
\newblock \bibinfo{publisher}{Lawrence Erlbaum}.

\bibitemdeclare{article}{zbi18}
\bibitem{zbi18}
\bibinfo{author}{Qinglin \surnamestart Zhang\surnameend},
  \bibinfo{author}{Chris \surnamestart Benton\surnameend} \&
  \bibinfo{author}{Daniela \surnamestart Inclezan\surnameend}
  (\bibinfo{year}{2019}): \emph{\bibinfo{title}{An Application of {ASP}
  Theories of Intentions to Understanding Restaurant Scenarios: Insights and
  Narrative Corpus}}.
\newblock {\sl \bibinfo{journal}{Theory and Practice of Logic Programming}},
  pp. \bibinfo{pages}{1--21}, \doi{10.1017/S1471068419000048}.

\bibitemdeclare{inproceedings}{zi17}
\bibitem{zi17}
\bibinfo{author}{Qinglin \surnamestart Zhang\surnameend} \&
  \bibinfo{author}{Daniela \surnamestart Inclezan\surnameend}
  (\bibinfo{year}{2017}): \emph{\bibinfo{title}{An Application of {ASP}
  Theories of Intentions to Understanding Restaurant Scenarios}}.
\newblock In: {\sl \bibinfo{booktitle}{Proceedings of PAoASP'17}}.

\end{thebibliography}
\end{document}